\documentclass[12pt]{article}
\usepackage{amsmath}
\usepackage{graphicx,psfrag,epsf}
\usepackage{enumerate}
\usepackage{natbib}
\usepackage{textcomp}
\usepackage[hyphens]{url} 
\usepackage{hyperref}
\providecommand{\tightlist}{%
  \setlength{\itemsep}{0pt}\setlength{\parskip}{0pt}}

\newcommand{\blind}{0}

\usepackage{mathtools,amssymb,booktabs,amsthm,todonotes,colortbl} 

\newtheorem{definition}{Definition}

\usepackage{mathptmx} \usepackage{caption}

\DeclareCaptionStyle{italic}{labelfont={bf},textfont={it},labelsep=colon}  

\usepackage{booktabs}
\usepackage{longtable}
\usepackage{array}
\usepackage{multirow}
\usepackage{wrapfig}
\usepackage{float}
\usepackage{colortbl}
\usepackage{pdflscape}
\usepackage{tabu}
\usepackage{threeparttable}
\usepackage{threeparttablex}
\usepackage[normalem]{ulem}
\usepackage{makecell}

\begin{document}

\def\spacingset#1{\renewcommand{\baselinestretch}%
{#1}\small\normalsize} \spacingset{1}


\if0\blind
{
  \title{\bf Visualizing probability distributions across bivariate cyclic temporal granularities}

  \author{
        Sayani Gupta \thanks{Email: \href{mailto:Sayani.Gupta@monash.edu}{\nolinkurl{Sayani.Gupta@monash.edu}}} \\
    Department of Econometrics and Business Statistics, Monash University, Australia\\
     and \\     Rob J Hyndman \\
    Department of Econometrics and Business Statistics, Monash University, Australia\\
     and \\     Dianne Cook \\
    Department of Econometrics and Business Statistics, Monash University, Australia\\
     and \\     Antony Unwin \\
    University of Augsburg, Germany\\
      }
  \maketitle
} \fi

\if1\blind
{
  \bigskip
  \bigskip
  \bigskip
  \begin{center}
    {\LARGE\bf Visualizing probability distributions across bivariate cyclic temporal granularities}
  \end{center}
  \medskip
} \fi

\bigskip
\begin{abstract}
Deconstructing a time index into time granularities can assist in exploration and automated analysis of large temporal data sets. This paper describes classes of time deconstructions using linear and cyclic time granularities. Linear granularities respect the linear progression of time such as hours, days, weeks and months. Cyclic granularities can be circular such as hour-of-the-day, quasi-circular such as day-of-the-month, and aperiodic such as public holidays. The hierarchical structure of granularities creates a nested ordering: hour-of-the-day and second-of-the-minute are single-order-up. Hour-of-the-week is multiple-order-up, because it passes over day-of-the-week. Methods are provided for creating all possible granularities for a time index. A recommendation algorithm provides an indication whether a pair of granularities can be meaningfully examined together (a ``harmony''), or when they cannot (a ``clash'').

Time granularities can be used to create data visualizations to explore for periodicities, associations and anomalies. The granularities form categorical variables (ordered or unordered) which induce groupings of the observations. Assuming a numeric response variable, the resulting graphics are then displays of distributions compared across combinations of categorical variables.

The methods implemented in the open source R package \texttt{gravitas} are consistent with a tidy workflow, with probability distributions examined using the range of graphics available in \texttt{ggplot2}.
\end{abstract}

\noindent%
\vfill

\newpage
\spacingset{1.45} 

\hypertarget{introduction}{%
\section{Introduction}\label{introduction}}

Temporal data are available at various resolutions depending on the context. Social and economic data are often collected and reported at coarse temporal scales such as monthly, quarterly or annually. With recent advancement in technology, more and more data are recorded at much finer temporal scales. Energy consumption may be collected every half an hour, energy supply may be collected every minute, and web search data might be recorded every second. As the frequency of data increases, the number of questions about the periodicity of the observed variable also increases. For example, data collected at an hourly scale can be analyzed using coarser temporal scales such as days, months or quarters. This approach requires deconstructing time in various possible ways called time granularities \citep{aigner2011visualization}.

It is important to be able to navigate through all of these time granularities to have multiple perspectives on the periodicity of the observed data. This aligns with the notion of EDA \citep{Tukey1977-jx} which emphasizes the use of multiple perspectives on data to help formulate hypotheses before proceeding to hypothesis testing. Visualizing probability distributions conditional on one or more granularities is an indispensable tool for exploration. Analysts are expected to comprehensively explore the many ways to view and consider temporal data. However, the plethora of choices and the lack of a systematic approach to do so quickly can make the task overwhelming.

Calendar-based graphics \citep{wang2020calendar} are useful in visualizing patterns in the weekly and monthly structure, and are helpful when checking for the effects of weekends or special days. Any temporal data at sub-daily resolution can also be displayed using this type of faceting \citep{Wickham2009pk} with days of the week, month of the year, or another sub-daily deconstruction of time. But calendar effects are not restricted to conventional day-of-week or month-of-year deconstructions. There can be many different time deconstructions, based on the calendar or on categorizations of time granularities.

Linear time granularities (such as hours, days, weeks and months) respect the linear progression of time and are non-repeating. One of the first attempts to characterize these granularities is due to \citet{Bettini1998-ed}. However, the definitions and rules defined are inadequate for describing non-linear granularities. Hence, there is a need to define some new time granularities, that can be useful in visualizations. Cyclic time granularities can be circular, quasi-circular or aperiodic. Examples of circular granularities are hour of the day and day of the week; an example of a quasi-circular granularity is day of the month; examples of aperiodic granularities are public holidays and school holidays.

Time deconstructions can also be based on the hierarchical structure of time. For example, hours are nested within days, days within weeks, weeks within months, and so on. Hence, it is possible to construct single-order-up granularities such as second of the minute, or multiple-order-up granularities such as second of the hour. The lubridate package \citep{Grolemund2011-vm} provides tools to access and manipulate common date-time objects. However, most of its accessor functions are limited to single-order-up granularities.

The motivation for this work stems from the desire to provide methods to better understand large quantities of measurements on energy usage reported by smart meters in households across Australia, and indeed many parts of the world. Smart meters currently provide half-hourly use in kWh for each household, from the time they were installed, some as early as 2012. Households are distributed geographically and have different demographic properties as well as physical properties such as the existence of solar panels, central heating or air conditioning. The behavioral patterns in households vary substantially; for example, some families use a dryer for their clothes while others hang them on a line, and some households might consist of night owls, while others are morning larks. It is common to see aggregates \citep[see][]{Goodwin_2012} of usage across households, such as half-hourly total usage by state, because energy companies need to plan for maximum loads on the network. But studying overall energy use hides the distribution of usage at finer scales, and makes it more difficult to find solutions to improve energy efficiency. We propose that the analysis of smart meter data will benefit from systematically exploring energy consumption by visualizing the probability distributions across different deconstructions of time to find regular patterns and anomalies. Although we were motivated by the smart meter example, the problem and the solutions we propose are practically relevant to any temporal data observed more than once per year. In a broader sense, it could be even suitable for data observed by years, decades, and centuries as might be in weather or astronomical data.

This work provides tools for systematically exploring bivariate granularities within the tidy workflow (\citet{Grolemund2018-po}). In particular, we

\begin{itemize}
\tightlist
\item
  provide a formal characterization of cyclic granularities;
\item
  facilitate manipulation of single- and multiple-order-up time granularities through cyclic calendar algebra;
\item
  develop an approach to check the feasibility of creating plots or drawing inferences for any two cyclic granularities.
\end{itemize}

The remainder of the paper is organized as follows: Section~\ref{sec:linear-time} provides some background material on linear granularities and calendar algebra for computing different linear granularities. Section~\ref{sec:cyclic-gran} formally characterizes different cyclic time granularities by extending the framework of linear time granularities, and introducing cyclic calendar algebra for computing cyclic time granularities. The data structure for exploring the conditional distributions of the associated time series across pairs of cyclic time granularities is discussed in Section~\ref{sec:data-structure}. Section~\ref{sec:visualization} discusses the role of different factors in constructing an informative and trustworthy visualization. Section~\ref{sec:application} examines how systematic exploration can be carried out for a temporal and non-temporal application. Finally, we summarize our results and discuss possible future directions in Section~\ref{sec:discussion}.

\hypertarget{sec:linear-time}{%
\section{Linear time granularities}\label{sec:linear-time}}

Discrete abstractions of time such as weeks, months or holidays can be thought of as ``time granularities''. Time granularities are \textbf{linear} if they respect the linear progression of time. There have been several attempts to provide a framework for formally characterizing time granularities, including \citet{Bettini1998-ed} which forms the basis of the work described here.

\hypertarget{definitions}{%
\subsection{Definitions}\label{definitions}}

\begin{definition}\label{def:definition}
A \textbf{time domain} is a pair $(T; \le)$ where $T$ is a non-empty set of time instants and $\le$ is a total order on $T$.
\end{definition}

\noindent The time domain is assumed to be \emph{discrete}, and there is unique predecessor and successor for every element in the time domain except for the first and last.

\begin{definition}\label{def:index set}
The \textbf{index set}, $Z=\{z: z \in \mathbb{Z}_{\geq 0}\}$, uniquely maps the time instants to the set of non-negative integers.
\end{definition}

\begin{definition}\label{def:linear}
A \textbf{linear granularity} is a mapping $G$ from the index set, $Z$, to subsets of the time domain such that:
  (1) if $i < j$ and $G(i)$ and $G(j)$ are non-empty, then each element of $G(i)$ is less than all elements of $G(j)$; and
  (2) if $i < k < j$ and $G(i)$ and $G(j)$ are non-empty, then $G(k)$ is non-empty.
Each non-empty subset $G(i)$ is called a \textbf{granule}.
\end{definition}

\begin{figure}[!htb]

{\centering \includegraphics[width=\textwidth]{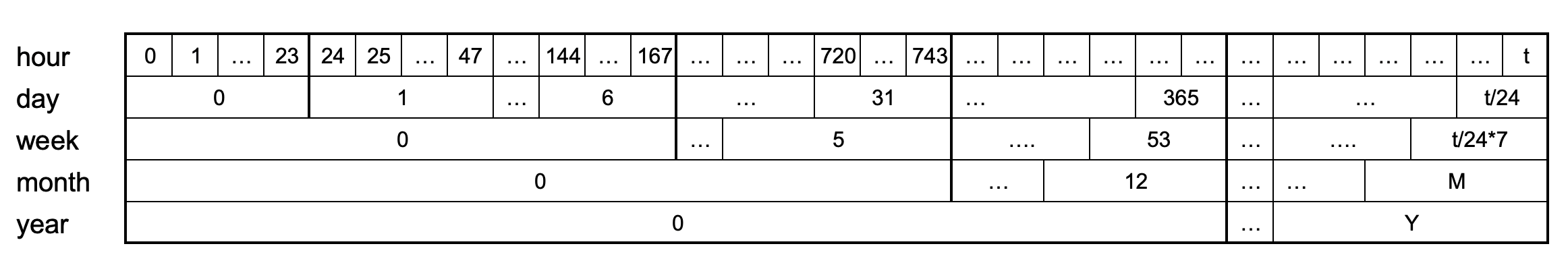} 

}

\caption{Illustration of time domain, linear granularities and index set. Hour, day, week, month and year are linear granularities and can also be considered to be time domains. These are ordered with ordering guided by integers and hence is unidirectional and non-repeating. Hours could also be considered the index set, and a bottom granularity.}\label{fig:linear-time}
\end{figure}

\noindent This implies that the granules in a linear granularity are non-overlapping, continuous and ordered. The indexing for each granule can also be associated with a textual representation, called the label. A discrete time model often uses a fixed smallest linear granularity named by \citet{Bettini1998-ed} \textbf{bottom granularity}. \autoref{fig:linear-time} illustrates some common linear time granularities. Here, ``hour'' is the bottom granularity and ``day'', ``week'', ``month'' and ``year'' are linear granularities formed by mapping the index set to subsets of the hourly time domain. If we have ``hour'' running from \(\{0, 1, \dots,t\}\), we will have ``day'' running from \(\{0, 1, \dots, \lfloor t/24\rfloor\}\). These linear granularities are uni-directional and non-repeating.

\hypertarget{relativities}{%
\subsection{Relativities}\label{relativities}}

Properties of pairs of granularities fall into various categories.

\begin{definition}\label{def:finerthan}
A linear granularity $G$ is \textbf{finer than} a linear granularity $H$, denoted $G \preceq H$, if for each index $i$, there exists an index $j$ such that
$G(i) \subset H(j).$
\end{definition}

\begin{definition}\label{def:groupsinto}
A linear granularity $G$ \textbf{groups into} a linear granularity $H$, denoted
$G \trianglelefteq H$, if for each index $j$ there exists a (possibly infinite) subset $S$ of the integers such that $H(j) = \bigcup_{i \in S}G(i).$
\end{definition}

\noindent For example, both \(day \trianglelefteq week\) and \(day \preceq week\) hold, since every granule of \(week\) is the union of some set of granules of day and each day is a subset of a \(week\). The relationship has period 7.

The relationship \(day \trianglelefteq month\) has a more complicated period. If leap years are ignored, each month is a grouping of the same number of days over years, hence the period of the grouping \((day, month)\) is one year. With the inclusion of leap years, the grouping period is 400 years.

\begin{definition}\label{def:periodical}
A granularity $G$ is \textbf{periodical} with respect to a granularity $H$ if:
(1) $G \trianglelefteq H$; and
(2) there exist $R$, $P \in \mathbb{Z}_+$, where $R$ is less than the number of granules of $H$, such that for all $i \in \mathbb{Z}$, if $H(i) = \bigcup_{j \in S}G(j)$ and $H (i + R) \neq \phi$ then $H (i + R) = \bigcup_{j \in S} G(j + P)$.
\end{definition}

For example, day is periodical with respect to week with \(R=1\) and \(P=7\), while (if we ignore leap years) day is periodical with respect to month with \(R=12\) and \(P=365\).

Granularities can also be periodical with respect to other granularities, except for a finite number of periods where they behave in an anomalous way; these are called \textbf{quasi-periodic} relationships \citep{Bettini2000-vy}. In a Gregorian calendar with leap years, day groups quasi-periodically into month with the exceptions of the time domain corresponding to \(29^{\text{th}}\) February of any year.

\begin{definition}\label{def:order}
The \textbf{order} of a linear granularity is the level of coarseness associated with a linear granularity. A linear granularity G will have lower order than H if each granule of G is composed of lower number of granules of bottom granularity than each granule of H.
\end{definition}

With two linear granularities \(G\) and \(H\), if \(G\) \emph{groups into} or \emph{finer than} \(H\) then \(G\) is of lower order than \(H\). For example, if the bottom granularity is day, then granularity week will have lower order than month since each week consist of fewer days than each month.

Granules in any granularity may be aggregated to form a coarser granularity. A system of multiple granularities in lattice structures is referred to as a \textbf{calendar} by \citet{Dyreson_2000}. Linear time granularities are computed through ``calendar algebra'' operations \citep{Ning_2002} designed to generate new granularities recursively from the bottom granularity. For example, due to the constant length of day and week, we can derive them from hour using
\[
  D(j) = \lfloor H(i)/24\rfloor, \qquad W(k) = \lfloor H(i)/(24*7)\rfloor,
\]
where \(H\), \(D\) and \(W\) denote hours, days and weeks respectively.

\hypertarget{sec:cyclic-gran}{%
\section{Cyclic time granularities}\label{sec:cyclic-gran}}

Cyclic granularities represent cyclical repetitions in time. They can be thought of as additional categorizations of time that are not linear. Cyclic granularities can be constructed from two linear granularities, that relate periodically; the resulting cycles can be either \emph{regular} (\textbf{circular}), or \emph{irregular} (\textbf{quasi-circular}).

\hypertarget{sec:circular-gran-def}{%
\subsection{Circular granularities}\label{sec:circular-gran-def}}

\begin{definition}\label{def:circular}
A \textbf{circular granularity} $C_{B, G}$ relates linear granularity $G$ to bottom granularity $B$ if
\begin{equation} \label{eq:circular-gran}
\begin{split}
C_{B, G}(z) & = z~\text{mod}~P(B, G) \quad \forall z \in \mathbb{Z}_{\geq 0} \\
\end{split}
\end{equation}
where
$z$ denotes the index set,
$B$ groups periodically into $G$ with regular mapping and period $P(B, G)$.
\end{definition}

\begin{figure}[!htb]

{\centering \includegraphics[width=\textwidth]{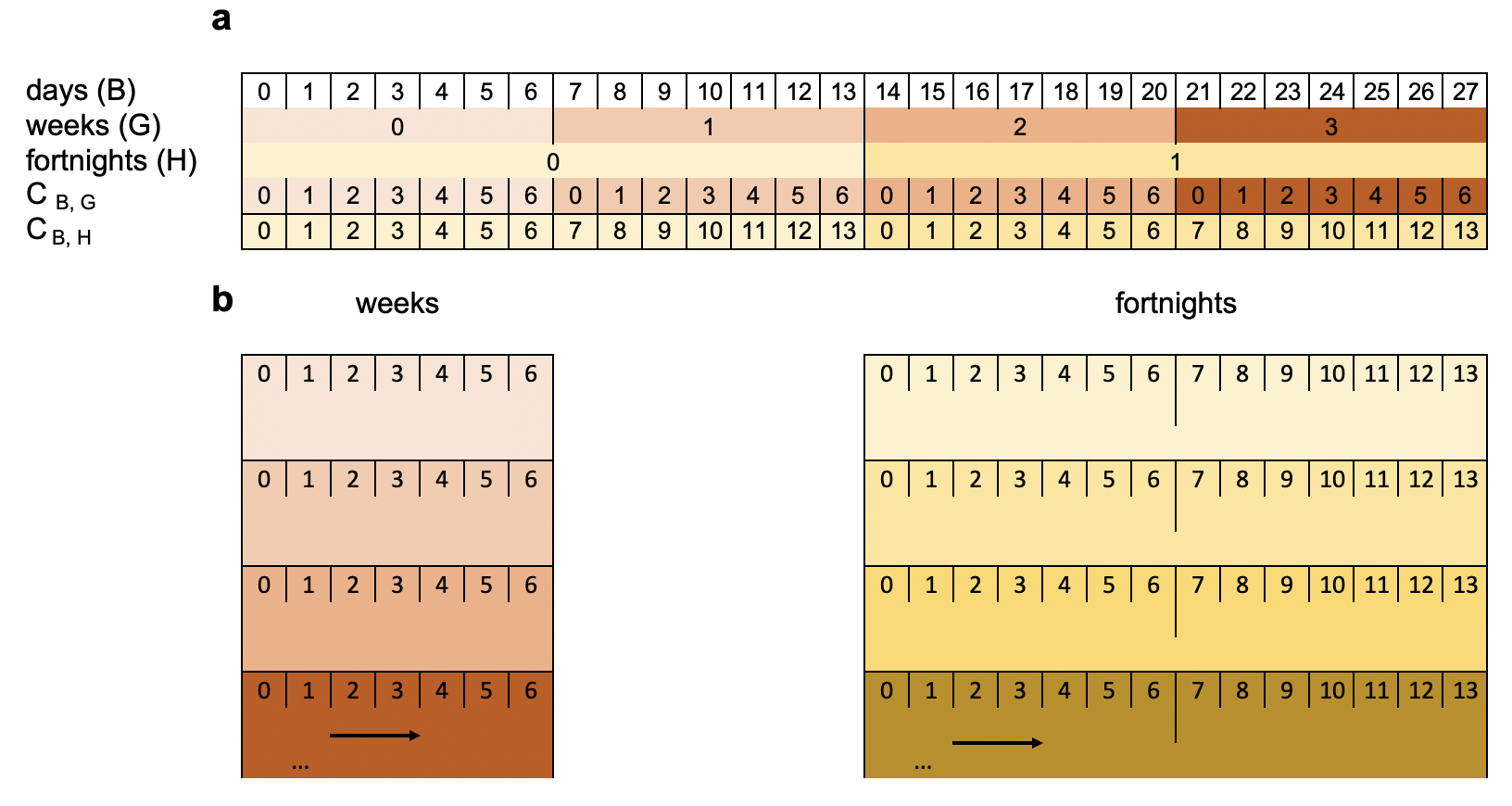} 

}

\caption{a. Index sets for some linear and circular granularities. b. Circular granularities can be constructed by slicing the linear granularity into pieces and stacking them.}\label{fig:circular-dow}
\end{figure}

\autoref{fig:circular-dow} illustrates some linear and cyclical granularities. Cyclical granularities are constructed be cutting the linear granularity into pieces, and stacking them to match the cycles (as shown in b). \(B, G, H\) (day, week, fortnight, respectively) are linear granularities. The circular granularity \(C_{B, G}\) (day-of-week) is constructed from \(B\) and \(G\), while circular granularity \(C_{B, H}\) (day-of-fortnight) is constructed from \(B\) and \(H\). These overlapping cyclical granularities share elements from the linear granularity. Each of \(C_{B , G}\) and \(C_{B , H}\) consist of repeated patterns \(\{0, 1, \dots, 6\}\) and \(\{0, 1, \dots, 13\}\) with \(P=7\) and \(P=14\) respectively.

Suppose \({L}\) is a label mapping that defines a unique label for each index \(\ell \in \{ 0,1,\dots, (P-1)\}\). For example, the label mapping \(L\) for \(C_{B, G}\) can be defined as
\[
  L: \{0,1, \dots, 6\} \longmapsto\ \{\text{Sunday}, \text{Monday}, \dots, \text{Saturday}\}.
\]

In general, any circular granularity relating two linear granularities can be expressed as
\[
  C_{(G, H)}(z) = \lfloor z/P(B,G) \rfloor~\text{mod}~P(G,H),
\]
where \(H\) is periodic with respect to \(G\) with regular mapping and period \(P(G,H)\). Table~\ref{tab:definitions} shows several circular granularities constructed using minutes as the bottom granularity.

\begin{table}[ht]
\begin{center}
\begin{tabular}{lll}
\toprule
Circular granularity & Expression & Period \\
\midrule
minute-of-hour                               &
  $C_1 = z ~\text{mod}~60$                     &
  $P_1 = \phantom{99}60$ \\
minute-of-day                                &
  $C_j = z ~\text{mod}~60*24$                  &
  $P_2= 1440$\\
hour-of-day                                  &
  $C_3 = \lfloor z/60\rfloor~\text{mod}~24$    &
  $P_3 = \phantom{99}24$ \\
hour-of-week                                 &
  $C_4 = \lfloor z/60\rfloor~\text{mod}~24*7$  &
  $P_4= \phantom{9}168$\\
day-of-week                                  &
  $C_5 = \lfloor z/24*60\rfloor ~\text{mod}~7$ &
  $P_5= \phantom{999}7$\\
\bottomrule
\end{tabular}
\end{center}
\caption{Examples of circular granularities with bottom granularity minutes. Circular granularity $C_i$ relates two linear granularities one of which groups periodically into the other with regular mapping and period $P_i$. Circular granularities can be expressed using modular arithmetic due to their regular mapping. }
\label{tab:definitions}
\end{table}

\hypertarget{sec:quasi-circular-gran-def}{%
\subsection{Quasi-circular granularities}\label{sec:quasi-circular-gran-def}}

A \textbf{quasi-circular} granularity cannot be defined using modular arithmetic because of the irregular mapping. However, they are still formed with linear granularities, one of which groups periodically into the other. \autoref{tab:quasi} shows some examples of quasi-circular granularities.

\begin{table}[ht]
\centering
\begin{tabular}{lr@{~}lr@{~}r}
\toprule
Quasi-circular granularity && Possible period lengths\\
\midrule
$Q_1 =$ day-of-month && $P_1 = 31, 30, 29, 28$\\
$Q_2 =$ hour-of-month && $P_2 = 24\times 31, 24\times 30, 24\times 29, 24\times 28$\\
$Q_3 =$ day-of-year && $P_3 = 366, 365$\\
$Q_4 =$ week-of-month && $P_4 = 5, 4$\\
\bottomrule
\end{tabular}
\caption{Examples of quasi-circular granularities relating two linear granularities with irregular mapping leading to several possible period lengths.}
\label{tab:quasi}
\end{table}

\begin{definition}\label{def:quasicircular}
A \textbf{quasi-circular granularity} $Q_{B, G'}$ is formed when bottom granularity $B$ groups periodically into linear granularity $G'$ with irregular mapping such that the granularities are given by
\begin{equation}\label{eq:quasi}
Q_{B, G'}(z) =
z - \sum_{w=0}^{k-1}\vert T_{w ~\text{mod}~R'} \vert, \quad \text{for}\quad z \in T_{k},
\end{equation}
where
$z$ denotes the index set,
$R'$ is the number of granules of $G'$ in each repetition of the grouping,
$T_w$ are the sets of indices of $B$ such that $G'(w) = \bigcup_{z \in T_w}B(z)$,
and $\vert T_w \vert$ is the cardinality of set $T_w$.
\end{definition}

For example, day-of-year is quasi-periodic with either 365 or 366 granules of \(B\) (days) within each period of \(G'\) (years). The pattern repeats every \(R'=400\) years. So \(Q_{B, G'}\) is a repetitive categorization of time, similar to circular granularities, except that the number of granules of \(B\) is not the same across different granules of \(G'\).

\hypertarget{sec:aperiodic-gran-def}{%
\subsection{Aperiodic granularities}\label{sec:aperiodic-gran-def}}

Aperiodic time granularities are those that cannot be specified as a periodic repetition of a pattern of granules. Most public holidays repeat every year, but there is no reasonably small period within which their behavior remains constant. A classic example is Easter (in the western tradition) whose dates repeat only after 5.7 million years \citep{Reingold2001-kf}. In Australia, if a standard public holiday falls on a weekend, a substitute public holiday will sometimes be observed on the first non-weekend day (usually Monday) after the weekend. Examples of aperiodic granularity may also include school holidays or a scheduled event. All of these are recurring events, but with non-periodic patterns. Consequently, \(P_i\) (as given in \autoref{tab:quasi}) are essentially infinite for aperiodic granularities.

\begin{definition}\label{def:aperiodic}
An \textbf{aperiodic cyclic granularity} is formed when bottom granularity $B$ groups aperiodically into linear granularity $M$ such that the granularities are given by
\begin{equation}\label{eq:aperiodic}
A_{B, M}(z) = \begin{cases}
                  i, & \text{for}\quad z \in T_{i_j} \\
                  0  & \text{otherwise},
                \end{cases}
\end{equation}
where
$z$ denotes the index set,
$T_{i_j}$ are the sets of indices of $B$ describing aperiodic linear granularities $M_{i}$ such that $M_{i}(j) = \bigcup_{z \in T_{i_j}}B(z)$, and $M = \bigcup_{i=1}^{n}M_{i}$.
\end{definition}

\begin{figure}[!htb]

{\centering \includegraphics[width=\textwidth]{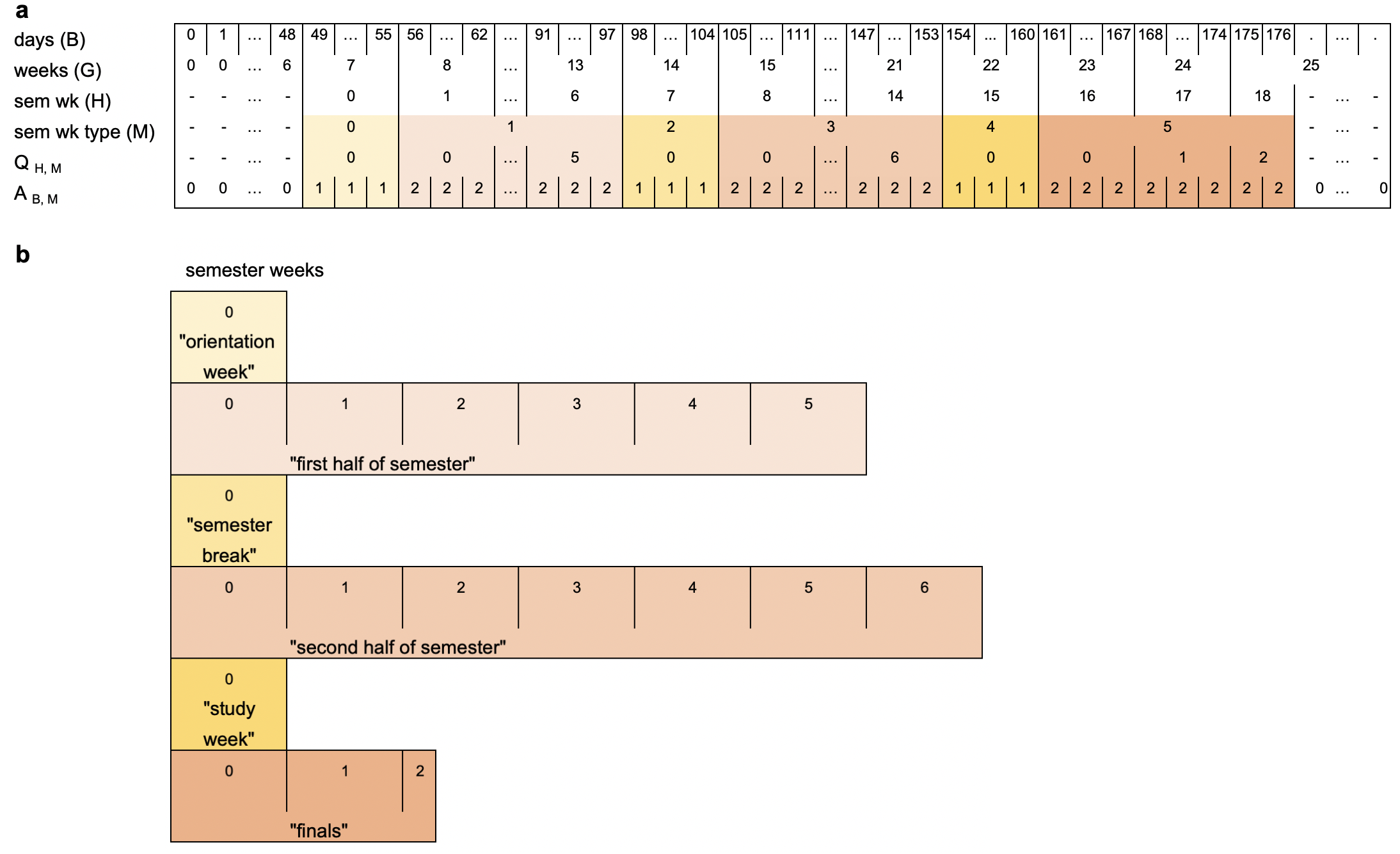} 

}

\caption{Quasi-circular and aperiodic cyclic granularities illustrated through (a) linear and (b) stacked displays of time. The linear display shows granularities days, weeks, semester weeks, semester week type distributed over linear time. Here a semester lasts for 18 weeks and 2 days, each starting with one week of orientation followed by an in-session period of 6 weeks, a semester-break of 1 week, an in-session period of 7 weeks, a 1-week study break, before final exams which continue for 16 days. This pattern remains same for all semester and hence \(Q_{H, M}\) with \(P\) = 128 days will be a quasi-circular granularity with repeating patterns. \(A_{B, M}\) will be an aperiodic cyclic granularity since the placement of the semester within an year varies across years.}\label{fig:aperiodic-example}
\end{figure}

For example, consider the school semester shown in \autoref{fig:aperiodic-example}. Let the linear granularities \(M_1\) and \(M_2\) denote the in-session semester period and semester break period respectively. Both \(M_1\), \(M_2\) and \(M = M_{1}\bigcup M_{2}\) denoting the ``semester week type'' are aperiodic with respect to days (\(B\)) or weeks (\(G\)). Hence \(A_{B, M}\) denoting day-of-the-``semester week type'' would be an aperiodic cyclic granularity, because the placement of the semester within an year would vary across years. Here, \(Q_{H, M}\) denoting week-of-the-``semester week type'' would be a quasi-circular granularity since the distribution of semester weeks within a semester is assumed to remain constant over years.

\hypertarget{sec:cyclic-calendar}{%
\subsection{Relativities}\label{sec:cyclic-calendar}}

The hierarchical structure of time creates a natural nested ordering which can be used in the computation of relative pairs of granularities.

\begin{definition}\label{def:hierarchy}
The nested ordering of linear granularities can be organized into a \textbf{hierarchy table}, denoted as $H_n: (G, C, K)$, which arranges them from lowest to highest in order. It shows how the $n$ granularities relate through $K$, and how the cyclic granularities, $C$, can be defined relative to the linear granularities. Let $G_{\ell}$ and $G_{m}$ represent the linear granularity of order $\ell$ and $m$ respectively with $\ell<m$. Then $K \equiv P(\ell,m)$ represents the period length of the grouping $(G_{\ell}, G_{m})$, if $C_{G_{\ell}, G_{m}}$ is a circular granularity and $K \equiv k(\ell,m)$ represents the operation to obtain $G_{m}$ from $G_{\ell}$, if $C_{G_{\ell}, G_{m}}$ is quasi-circular.
\end{definition}

For example, \autoref{tab:tab-mayan} shows the hierarchy table for the Mayan calendar. In the Mayan calendar, one day was referred to as a kin and the calendar was structured such that 1 kin = 1 day; 1 uinal = 20 kin; 1 tun = 18 uinal (about a year); 1 katun = 20 tun (20 years) and 1 baktun = 20 katun.

\begin{table}

\caption{\label{tab:tab-mayan}Hierarchy table for Mayan calendar with circular single-order-up granularities.}
\centering
\begin{tabular}[t]{lll}
\toprule
linear (G) & single-order-up cyclic (C) & period length/conversion operator (K)\\
\midrule
kin & kin-of-uinal & 20\\
uinal & uinal-of-tun & 18\\
tun & tun-of-katun & 20\\
katun & katun-of-baktun & 20\\
baktun & 1 & 1\\
\bottomrule
\end{tabular}
\end{table}

Like most calendars, the Mayan calendar used the day as the basic unit of time \citep{Reingold2001-kf}. The structuring of larger units, weeks, months, years and cycle of years, though, varies substantially between calendars. For example, the French revolutionary calendar divided each day into 10 ``hours'', each ``hour'' into 100 ``minutes'' and each ``minute'' into 100 ``seconds'', the duration of which is 0.864 common seconds. Nevertheless, for any calendar a hierarchy table can be defined. Note that it is not always possible to organize an aperiodic linear granularity in a hierarchy table. Hence, we assume that the hierarchy table consists of periodic linear granularities only, and that the cyclic granularity \(C_{G(\ell),G(m)}\) is either circular or quasi-circular.

\begin{definition}\label{def:norderup}
The hierarchy table contains \textbf{multiple-order-up} granularities which are cyclic granularities that are nested within multiple levels.
A \textbf{single-order-up} is a cyclic granularity which is nested within a single level. It is a special case of multiple-order-up granularity.
\end{definition}

\noindent In the Mayan calendar (Table \ref{tab:tab-mayan}), kin-of-tun or kin-of-baktun are examples of multiple-order-up granularities and single-order-up granularities are kin-of-uinal, uinal-of-tun etc.

\hypertarget{computation}{%
\subsection{Computation}\label{computation}}

Following the calendar algebra of \citet{Ning_2002} for linear granularities, we can define cyclic calendar algebra to compute cyclic granularities. Cyclic calendar algebra comprises two kinds of operations:
(1) \textbf{single-to-multiple} (the calculation of \emph{multiple-order-up} cyclic granularities from \emph{single-order-up} cyclic granularities) and (2) \textbf{multiple-to-single} (the reverse).

\hypertarget{single-to-multiple-order-up}{%
\subsubsection*{Single-to-multiple order-up}\label{single-to-multiple-order-up}}
\addcontentsline{toc}{subsubsection}{Single-to-multiple order-up}

Methods to obtain multiple-order-up granularity will depend on whether the hierarchy consists of all circular single-order-up granularities or a mix of circular and quasi-circular single-order-up granularities. Circular single-order-up granularities can be used recursively to obtain a multiple-order-up circular granularity using
\begin{equation} \label{eq:eq7}
C_{G_\ell,G_m}(z)
  = \sum_{i=0}^{m - \ell - 1} P(\ell, \ell+i)C_{G_{\ell+i},G_{\ell+i+1}}(z),
\end{equation}
where \(\ell < m - 1\) and \(P(i, i) = 1\) for \(i=0,1,\dots,m-\ell-1\), and
\(C_{B, G}(z) = z~\text{mod}~P(B, G)\) as per Equation \eqref{eq:circular-gran}.

For example, the multiple-order-up granularity \(C_{\text{uinal},\text{katun}}\) for the Mayan calendar could be obtained using
\begin{align*}
C_{\text{uinal}, \text{baktun}}(z)  &=  C_{\text{uinal}, \text{tun}}(z) + P(\text{uinal}, \text{tun})C_{\text{tun},\text{katun}}(z) + P(\text{uinal},\text{katun})C_{\text{katun}, \text{baktun}}(z) \\
               &=  \lfloor z/20\rfloor ~\text{mod}~18 + 18\lfloor 18\times z/20\rfloor ~\text{mod}~20
                + 18\times20\lfloor 18\times20\times z/20\rfloor ~\text{mod}~20.
\end{align*}

Now consider the case where there is one quasi-circular single order-up granularity in the hierarchy table while computing a multiple-order-up quasi-circular granularity. Any multiple-order-up quasi-circular granularity \(C_{\ell, m}(z)\) could then be obtained as a discrete combination of circular and quasi-circular granularities.

Depending on the order of the combination, two different approaches need to be employed leading to the following cases:

\begin{itemize}
\item
  \(C_{\ell,m'}(z)\) is circular and \(C_{m',m}(z)\) is quasi-circular
  \begin{equation} \label{eq:multifromsingle-quasi1}
  C_{G_\ell,G_{m}}(z) = C_{G_{\ell},G_{m'}}(z) + P(\ell, m')C_{G_{m'},G_{m}}(z)
  \end{equation}
\item
  \(C_{\ell,m'}(z)\) is quasi-circular and \(C_{m',m}(z)\) is circular
  \begin{equation} \label{eq:multifromsingle-quasi2}
  C_{G_\ell,G_{m}}(z)  = C_{G_{\ell},G_{m'}}(z) + \sum_{w=0}^{C_{m',m}(z) -1}(\vert T_{w} \vert)
  \end{equation}
  where, \(T_w\) is such that \(G_{m'}(w) = \bigcup_{z \in T_w}G_{\ell}\) and \(\vert T_w \vert\) is the cardinality of set \(T_w\).
\end{itemize}

\begin{table}

\caption{\label{tab:tab-gregorian}Hierarchy table for the Gregorian calendar with both circular and quasi-circular single-order-up granularities.}
\centering
\begin{tabular}[t]{lll}
\toprule
linear (G) & single-order-up cyclic (C) & period length/conversion operator (K)\\
\midrule
minute & minute-of-hour & 60\\
hour & hour-of-day & 24\\
day & day-of-month & k(day, month)\\
month & month-of-year & 12\\
year & 1 & 1\\
\bottomrule
\end{tabular}
\end{table}

For example, the Gregorian calendar (\autoref{tab:tab-gregorian}) has day-of-month as a single-order-up quasi-circular granularity, with the other granularities being circular. Using Equations \eqref{eq:multifromsingle-quasi1} and \eqref{eq:multifromsingle-quasi2}, we then have:
\[
  C_{hour, month}(z) = C_{hour, day}(z) + P(hour, day)*C_{day, month}(z)
\]
\[
  C_{day, year}(z) = C_{day,month}(z) + \sum_{w=0}^{C_{month, year}(z)-1}(\vert T_{w} \vert),
\]
where \(T_w\) is such that \(month(w) = \bigcup_{z \in T_w}day(z)\).

\hypertarget{multiple-to-single-order-up}{%
\subsubsection*{Multiple-to-single order-up}\label{multiple-to-single-order-up}}
\addcontentsline{toc}{subsubsection}{Multiple-to-single order-up}

Similar to single-to-multiple operations, multiple-to-single operations involve different approaches for all circular single-order-up granularities and a mix of circular and quasi-circular single-order-up granularities in the hierarchy. For a hierarchy table \(H_n: (G, C, K)\) with only circular single-order-up granularities and \(\ell_1, \ell_2, m_1, m_2 \in {1, 2, \dots, n}\) and \(\ell_2<\ell_1\) and \(m_2>m_1\), multiple-order-up granularities can be obtained using \eqref{eq:all-circular-multiple}.
\begin{equation} \label{eq:all-circular-multiple}
C_{G_{\ell_1}, G_{m_1}}(z) = \lfloor C_{G_{\ell_2}, G_{m_2}}(z)/P(\ell_2,\ell_1) \rfloor ~\text{mod}~P(\ell_1, m_1)
\end{equation}
For example, in the Mayan Calendar, it is possible to compute the single-order-up granularity tun-of-katun from uinal-of-baktun, since \(C_{tun, katun}(z) = \lfloor C_{uinal, baktun}(z)/18\rfloor ~\text{mod}~20\).

\hypertarget{multiple-order-up-quasi-circular-granularities}{%
\subsubsection*{Multiple order-up quasi-circular granularities}\label{multiple-order-up-quasi-circular-granularities}}
\addcontentsline{toc}{subsubsection}{Multiple order-up quasi-circular granularities}

Single-order-up quasi-circular granularity can be obtained from multiple-order-up quasi-circular granularity and single/multiple-order-up circular granularity using Equations \eqref{eq:multifromsingle-quasi1} and \eqref{eq:multifromsingle-quasi2}.

\hypertarget{sec:data-structure}{%
\section{Data structure}\label{sec:data-structure}}

Effective exploration and visualization benefits from well-organized data structures. \citet{wang2020tsibble} introduced the tidy ``tsibble'' data structure to support exploration and modeling of temporal data. This forms the basis of the structure for cyclic granularities. A tsibble comprises an index, optional key(s), and measured variables. An index is a variable with inherent ordering from past to present and a key is a set of variables that define observational units over time. A linear granularity is a mapping of the index set to subsets of the time domain. For example, if the index of a tsibble is days, then a linear granularity might be weeks, months or years. A bottom granularity is represented by the index of the tsibble.

\setlength{\aboverulesep}{0pt}\setlength{\belowrulesep}{0pt}

\begin{table}

\caption{\label{tab:data-structure}The data structure for exploring periodicities in data by including cyclic granularities in the tsibble structure with index, key and measured variables.}
\centering
\begin{tabular}[t]{lllllll}
\toprule
\cellcolor[HTML]{fdf2d0}{\textbf{index}} & \cellcolor[HTML]{fdf2d0}{\textbf{key}} & \cellcolor[HTML]{fdf2d0}{\textbf{measurements}} & \cellcolor[HTML]{fdf2d0}{\textbf{$C_1$}} & \cellcolor[HTML]{fdf2d0}{\textbf{$C_2$}} & \cellcolor[HTML]{fdf2d0}{\textbf{$\cdots$}} & \cellcolor[HTML]{fdf2d0}{\textbf{$C_{N_C}$}}\\
\midrule
\cellcolor[HTML]{dbe3f1}{ } & \cellcolor[HTML]{dbe3f1}{ } & \cellcolor[HTML]{dbe3f1}{ } & \cellcolor[HTML]{fdf2d0}{ } & \cellcolor[HTML]{fdf2d0}{ } & \cellcolor[HTML]{fdf2d0}{ } & \cellcolor[HTML]{fdf2d0}{ }\\
\bottomrule
\end{tabular}
\end{table}

\setlength{\aboverulesep}{0.4ex}\setlength{\belowrulesep}{0.65ex}

All cyclic granularities can be expressed in terms of the index set. \autoref{tab:data-structure} shows the tsibble structure (index, key, measurements) augmented by columns of cyclic granularities. The total number of cyclic granularities depends on the number of linear granularities considered in the hierarchy table and the presence of any aperiodic cyclic granularities. For example, if we have \(n\) periodic linear granularities in the hierarchy table, then \(n(n-1)/2\) circular or quasi-circular cyclic granularities can be constructed. Let \(N_C\) be the total number of contextual circular, quasi-circular and aperiodic cyclic granularities that can originate from the underlying periodic and aperiodic linear granularities. Simultaneously encoding more than a few of these cyclic granularities when visualizing the data overwhelms human comprehension. Instead, we focus on visualizing the data split by pairs of cyclic granularities (\(C_i\), \(C_j\)). Data sets of the form \textless{}\(C_i\), \(C_j\), \(v\)\textgreater{} then allow exploration and analysis of the measured variable \(v\).

\hypertarget{sec:synergy}{%
\subsection{Harmonies and clashes}\label{sec:synergy}}

The way granularities are related is important when we consider data visualizations. Consider two cyclic granularities \(C_i\) and \(C_j\), such that \(C_i\) maps index set to a set \(\{A_k \mid k=1,\dots,K\}\) and \(C_j\) maps index set to a set \(\{B_\ell \mid \ell =1,\dots,L\}\). Here, \(A_k\) and \(B_\ell\) are the levels/categories corresponding to \(C_i\) and \(C_j\) respectively. Let \(S_{k\ell}\) be a subset of the index set such that for all \(s \in S_{k\ell}\), \(C_i(s) = A_k\) and \(C_j(s) = B_\ell\). There are \(KL\) such data subsets, one for each combination of levels (\(A_k\), \(B_\ell\)). Some of these sets may be empty due to the structure of the calendar, or because of the duration and location of events in a calendar.

\begin{definition}\label{def:clash}
A \textbf{clash} is a pair of cyclic granularities that contains empty combinations of categories.
\end{definition}

\begin{definition}\label{def:harmony}
A \textbf{harmony} is a pair of cyclic granularities that does not contain any empty combinations of its categories.
\end{definition}

Structurally empty combinations can arise due to the structure of the calendar or hierarchy. For example, let \(C_i\) be day-of-month with 31 levels and \(C_j\) be week-of-month with 5 levels. There will be \(31\times 5=155\) sets \(S_{k\ell}\) corresponding to possible combinations of \(C_i\) and \(C_j\). Many of these are empty. For example, \(S_{1,5}\) is empty because the first day of the month can never correspond to the fifth week of the month. Hence the pair (day-of-month, week-of-month) is a clash.

Event-driven empty combinations arise due to differences in event location or duration in a calendar. For example, let \(C_i\) be day-of-week with 7 levels and \(C_j\) be working-day/non-working-day with 2 levels. While potentially all of these 14 sets \(S_{k\ell}\) can be non-empty (it is possible to have a public holiday on any day-of-week), in practice many of these will probably have very few observations. For example, there are few (if any) public holidays on Wednesdays or Thursdays in any given year in Melbourne, Australia.

An example of a harmony is where \(C_i\) and \(C_j\) denote day-of-week and month-of-year respectively. So \(C_i\) will have 7 levels while \(C_j\) will have 12 levels, giving \(12\times7=84\) sets \(S_{k\ell}\). All of these are non-empty because every day-of-week can occur in every month. Hence, the pair (day-of-week, month-of-year) is a harmony.

\hypertarget{sec:near-clashes}{%
\subsection{Near-clashes}\label{sec:near-clashes}}

Suppose \(C_i\) denotes day-of-year and \(C_j\) denotes day-of-week. While any day of the week can occur on any day of the year, some combinations will be very rare. For example, the 366th day of the year will only coincide with a Wednesday approximately every 28 years on average. We refer to these as ``near-clashes''.

\hypertarget{sec:visualization}{%
\section{Visualization}\label{sec:visualization}}

The grammar of graphics introduced a framework to construct statistical graphics by relating the data space to the graphic space \citep{Wilkinson1999-nk}. The layered grammar of graphics proposed by \citet{Wickham2009pk} gives an alternative and modified parametrization of the grammar, and suggests that graphics are made up of distinct layers of grammatical elements.

Drawing from the grammar of graphics, we consider visualizing the distribution of the measured variable \(v\) conditional on the values of two granularities, \(C_i\) and \(C_j\). The following layers can be specified:

\begin{itemize}
\tightlist
\item
  Data: \textless{}\(C_i\), \(C_j\), \(v\)\textgreater{};
\item
  Aesthetic mapping (mapping of variables to elements of the plot): \(C_i\) mapped to \(x\) position and \(v\) to \(y\) position;
\item
  Facet (split plots): \(C_j\);
\item
  Data summarization: any descriptive or smoothing statistics that summarizes the distribution of \(v\);
\item
  Geometric objects (physical representation of the data): any geometry displaying the distribution; for example, boxplot, letter value, violin, ridge or highest density region plots.
\end{itemize}

\hypertarget{data-summarization-and-geometric-objects}{%
\subsection{Data summarization and geometric objects}\label{data-summarization-and-geometric-objects}}

Plot selection is dictated by the choice of data summarization and geometric objects used for the visualization. The basic plot choice for our data structure is one that can display distributions using kernel density estimates or descriptive statistics. Displays based on descriptive statistics include variations of box plots \citep{Tukey1977-jx} such as notched box plots \citep{McGill1978-hg}, letter-value plots \citep{Hofmann2017-sg} or quantile plots. Plots based on kernel density estimates include violin plots \citep{Hintze1998-zi}, summary plot \citep{Potter2010-qc}, ridge line plots \citep{R-ggridges}, and highest density region (HDR) boxplots \citep{Hyndman1996-ft}. Each type of density display has parameters that need to be estimated from the data. Each has its own strengths and weaknesses that should be borne in mind while using them for exploration. Visualizing distributions can be uninformative or potentially misleading if data summarization are performed on rarely occurring categories (Section~\ref{sec:near-clashes}). Even when there are no rarely occurring events, the number of observations may vary greatly within or across each facet, due to missing observations or uneven locations of events in the time domain. In such cases, data summarization should be used with caution as sample sizes will directly affect the accuracy of the estimated quantities being displayed.

\hypertarget{sec:aesthetics}{%
\subsection{Facet and aesthetic variables}\label{sec:aesthetics}}

\hypertarget{levels}{%
\subsubsection*{Levels}\label{levels}}
\addcontentsline{toc}{subsubsection}{Levels}

The levels of cyclic granularities affect plotting choices since space and resolution may be problematic with too many levels. A potential approach could be to categorize the number of levels as low/medium/high/very high for each cyclic granularity and define some criteria based on human cognitive power, available display size and the aesthetic mappings. Default values for these categorizations could be chosen based on levels of common temporal granularities like days of the month, days of the fortnight, or days of the week.

\hypertarget{synergy-of-cyclic-granularities}{%
\subsubsection*{Synergy of cyclic granularities}\label{synergy-of-cyclic-granularities}}
\addcontentsline{toc}{subsubsection}{Synergy of cyclic granularities}

The synergy of the two cyclic granularities will affect plotting choices for exploratory analysis. Cyclic granularities that form clashes (Section \ref{sec:synergy}) or near-clashes lead to potentially ineffective graphs. Harmonies tend to be more useful for exploring patterns.

\begin{figure}[!htb]

\includegraphics[width=\textwidth]{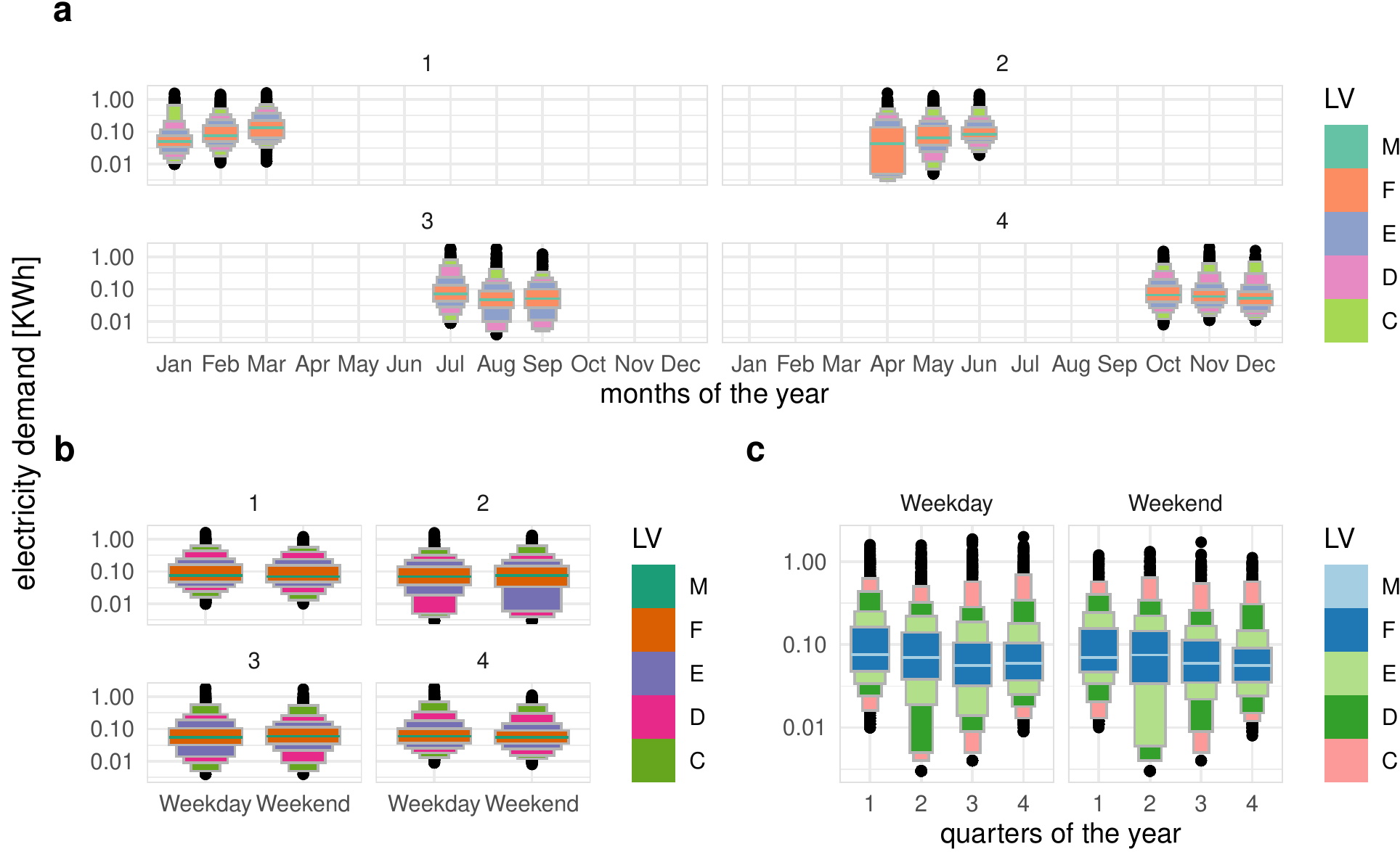} \hfill{}

\caption{Distribution of energy consumption displayed through letter value plots: plot a across month-of-year faceted by quarter-of-year; plot b across weekday/weekend faceted by quarter-of-year; plot c across quarter-of-year faceted by weekday/weekend. Plot a shows a clash since there are empty combinations. Plots b and c show harmonies since each quarter includes both weekdays and weekends. It can be seen in plot b that for every quarter (mapped to facet), weekend and weekday consumption are fairly similar except for the second quarter where the letter values below D and E behave differently. This is probably because of lower temperatures in the second quarter compared to the first quarter (summer in Australia). Plot c switches the granularities mapped to the x-axis and the facets and helps to compare quarters within weekdays and weekends. For example, for weekdays the interquartile range of consumption reduces over the year, whereas this pattern is not true for weekends.}\label{fig:allFig}
\end{figure}

\autoref{fig:allFig} (a) shows the distribution of half-hourly electricity consumption through letter value plots across months of the year faceted by quarters of the year. This plot does not work because quarter-of-the-year clashes with month-of-the-year, leading to empty subsets. For example, the first quarter never corresponds to December.

\hypertarget{interchangeability-of-mappings}{%
\subsubsection*{Interchangeability of mappings}\label{interchangeability-of-mappings}}
\addcontentsline{toc}{subsubsection}{Interchangeability of mappings}

When \(C_i\) is mapped to the \(x\) position and \(C_j\) to facets, then the \(A_k\) levels are juxtaposed and each \(B_\ell\) represent a group/facet. Gestalt theory suggests that when items are placed in close proximity, people assume that they are in the same group because they are close to one another and apart from other groups. Hence, in this case the \(A_k\)s are compared against each other within each group. With the mapping of \(C_i\) and \(C_j\) reversed, the emphasis will shift to comparing \(B_\ell\) levels rather than \(A_k\) levels.

For example, \autoref{fig:allFig} (b) shows the letter value plot across weekday/weekend faceted by quarters of the year and \autoref{fig:allFig} (c) shows the same two cyclic granularities with their mapping reversed. \autoref{fig:allFig} (b) helps us to compare weekday and weekend within each quarter and \autoref{fig:allFig} (c) helps to compare quarters within weekend and weekday.

\hypertarget{sec:application}{%
\section{Applications}\label{sec:application}}

\hypertarget{sec:smartmeter}{%
\subsection{Smart meter data of Australia}\label{sec:smartmeter}}

Smart meters provide large quantities of measurements on energy usage for households across Australia. One of the customer trials \citep{smart-meter} conducted as part of the Smart Grid Smart City project in Newcastle and parts of Sydney provides customer level data on energy consumption for every half hour from February 2012 to March 2014. We can use this data set to visualize the distribution of energy consumption across different cyclic granularities in a systematic way to identify different behavioral patterns.

\hypertarget{cyclic-granularities-search-and-computation}{%
\subsubsection*{Cyclic granularities search and computation}\label{cyclic-granularities-search-and-computation}}
\addcontentsline{toc}{subsubsection}{Cyclic granularities search and computation}

The tsibble object \texttt{smart\_meter10} from R package \texttt{gravitas} \citep{R-gravitas} includes the variables \texttt{reading\_datetime}, \texttt{customer\_id} and \texttt{general\_supply\_kwh} denoting the index, key and measured variable respectively. The interval of this tsibble is 30 minutes.

To identify the available cyclic time granularities, consider the conventional time deconstructions for a Gregorian calendar that can be formed from the 30-minute time index: half-hour, hour, day, week, month, quarter, half-year, year. In this example, we will consider the granularities hour, day, week and month giving six cyclic granularities ``hour\_day'', ``hour\_week'', ``hour\_month'', ``day\_week'', ``day\_month'' and ``week\_month'', read as ``hour of the day'', etc. To these we add day-type (``wknd\_wday'') to capture weekend and weekday behavior. Now that we have a list of cyclic granularities to look at, we can compute them using the results in Section~\ref{sec:cyclic-calendar}.

\hypertarget{screening-and-visualizing-harmonies}{%
\subsubsection*{Screening and visualizing harmonies}\label{screening-and-visualizing-harmonies}}
\addcontentsline{toc}{subsubsection}{Screening and visualizing harmonies}

Using these seven cyclic granularities, we want to explore patterns of energy behavior. Each of these seven cyclic granularities can either be mapped to the x-axis or to facets. Choosing \(2\) of the possible \(7\) granularities, gives \(^{7}P_2 = 42\) candidates for visualization. Harmonies can be identified among those \(42\) possibilities to narrow the search. \autoref{tab:harmony-tab} shows \(16\) harmony pairs after removing clashes and any cyclic granularities with more than \(31\) levels, as effective exploration becomes difficult with many levels (Section~\ref{sec:aesthetics}).

\begin{table}[!h]

\caption{\label{tab:harmony-tab}Harmonies with pairs of cyclic granularities, one mapped to facets and the other to the x-axis. Only 16 of 42 possible combinations of cyclic granularities are harmony pairs.}
\centering
\begin{tabular}[t]{llrr}
\toprule
facet variable & x-axis variable & facet levels & x-axis levels\\
\midrule
day\_week & hour\_day & 7 & 24\\
day\_month & hour\_day & 31 & 24\\
week\_month & hour\_day & 5 & 24\\
wknd\_wday & hour\_day & 2 & 24\\
hour\_day & day\_week & 24 & 7\\
\addlinespace
day\_month & day\_week & 31 & 7\\
week\_month & day\_week & 5 & 7\\
hour\_day & day\_month & 24 & 31\\
day\_week & day\_month & 7 & 31\\
wknd\_wday & day\_month & 2 & 31\\
\addlinespace
hour\_day & week\_month & 24 & 5\\
day\_week & week\_month & 7 & 5\\
wknd\_wday & week\_month & 2 & 5\\
hour\_day & wknd\_wday & 24 & 2\\
day\_month & wknd\_wday & 31 & 2\\
\addlinespace
week\_month & wknd\_wday & 5 & 2\\
\bottomrule
\end{tabular}
\end{table}

\begin{figure}[!htb]

{\centering \includegraphics[width=\textwidth]{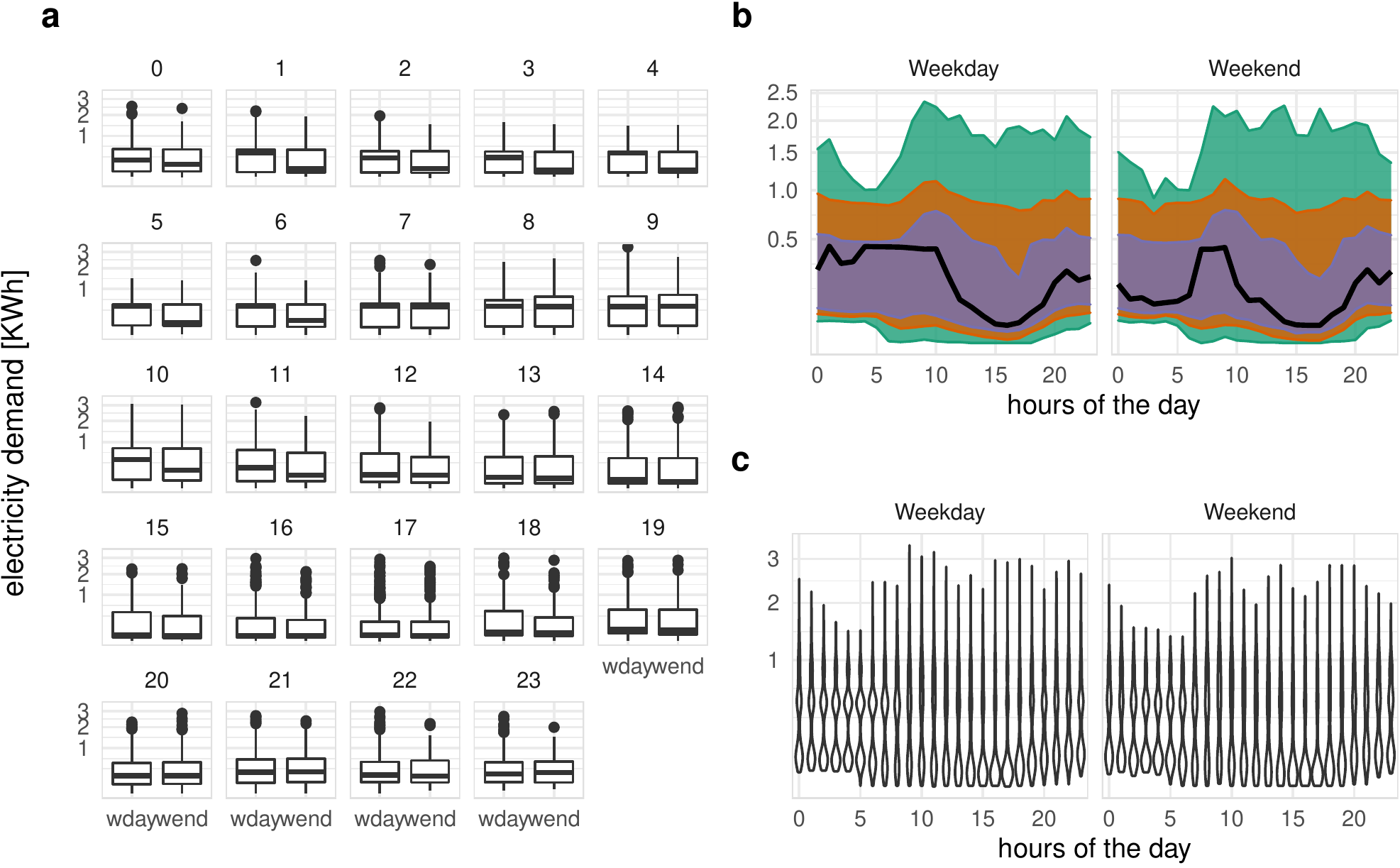} 

}

\caption{Energy consumption of a single customer shown with different distribution displays, and granularity arrangements. Two granularities are used: 1. hour of the day; and 2. weekday/weekend. Plot a shows granularity 1 faceted by granularity 2; plots b and c shows the reverse mapping. Plot a allows comparison of usage by workday within each hour of the day using side-by-side boxplots, showing that on work days there is more consumption early in the day. Plots b and c examine the temporal trend of consumption over the course of a day, separately for the type of day. Plot b uses an area quantile to emphasize the time component; e.g., median consumption shows prolonged high usage in the morning on weekdays. Plot c uses a violin plot emphasize subtler distributional differences across hours, showing that morning use on weekdays is bimodal, and on some work days there is low usage possibly indicating the person is working from home or having a late start.}\label{fig:bothcust}
\end{figure}

A few harmony pairs are displayed in \autoref{fig:bothcust} to illustrate the impact of different distribution plots and reverse mapping. For each of \autoref{fig:bothcust}b and c, \(C_i\) denotes day-type (weekday/weekend) and \(C_j\) is hour-of-day. The geometry used for displaying the distribution is chosen as area-quantiles and violins in \autoref{fig:bothcust}b and c respectively. \autoref{fig:bothcust}a shows the reverse mapping of \(C_i\) and \(C_j\) with \(C_i\) denoting hour-of-day and \(C_j\) denoting day-type with distribution geometrically displayed as boxplots.

In \autoref{fig:bothcust}b, the black line is the median, whereas the purple band covers the 25th to 75th percentile, the orange band covers the 10th to 90th percentile, and the green band covers the 1st to 99th percentile. The first facet represents the weekday behavior while the second facet displays the weekend behavior; energy consumption across each hour of the day is shown inside each facet. The energy consumption is extremely skewed with the 1st, 10th and 25th percentile lying relatively close whereas 75th, 90th and 99th lying further away from each other. This is common across both weekdays and weekends. For the first few hours on weekdays, median energy consumption starts and continues to be higher for longer compared to weekends.

The same data is shown using violin plots instead of quantile plots in \autoref{fig:bothcust}c. There is bimodality in the early hours of the day for weekdays and weekends. If we visualize the same data with reverse mapping of the cyclic granularities (\autoref{fig:bothcust}a), then the natural tendency would be to compare weekend and weekday behavior within each hour and not across hours. Then it can be seen that median energy consumption for the early morning hours is higher for weekdays than weekends. Also, outliers are more prominent in the latter hours of the day. All of these indicate that looking at different distribution geometry or changing the mapping can shed light on different aspects of energy behavior for the same sample.

\hypertarget{sec:cricket}{%
\subsection{T20 cricket data of Indian Premier League}\label{sec:cricket}}

Our proposed approach can be generalized to other hierarchical granularities where there is an underlying ordered index. We illustrate this with data from the sport cricket. Although there is no conventional time component in cricket, each ball can be thought to represent an ordering over the course of the game. In the Twenty20 format, an over will consist of 6 balls (with some exceptions), an innings is restricted to a maximum of 20 overs, a match will consist of 2 innings and a season consists of several matches. Thus, there is a hierarchy where ball is nested within overs, overs nested within innings, and innings within matches. Cyclic granularities can be constructed using this hierarchy. Example granularities include ball of the over, over of the innings, and ball of the innings. The hierarchy table is given in \autoref{tab:hierarchy-cric}.

\begin{table}

\caption{\label{tab:hierarchy-cric}Hierarchy table for cricket where overs are nested within an innings, innings nested within a match and matches within a season.}
\centering
\begin{tabular}[t]{lll}
\toprule
linear (G) & single-order-up cyclic (C) & period length/conversion operator (K)\\
\midrule
over & over-of-inning & 20\\
inning & inning-of-match & 2\\
match & match-of-season & k(match, season)\\
season & 1 & 1\\
\bottomrule
\end{tabular}
\end{table}

Although most of these cyclic granularities are circular by design of the hierarchy, in practice some granularities are aperiodic. For example, most overs will consist of 6 balls, but there are exceptions due to wide balls, no-balls, or when an innings finishes before the over finishes. Thus, the cyclic granularity ball-of-over may be aperiodic.

The Indian Premier League (IPL) is a professional Twenty20 cricket league in India contested by eight teams representing eight different cities in India. The IPL ball-by-ball data is provided in the \texttt{cricket} data set in the \texttt{gravitas} package for a sample of 214 matches spanning 9 seasons (2008 to 2016) such that each over has 6 balls, each innings has 20 overs and each match has 2 innings.

\begin{figure}

{\centering \includegraphics[width=\textwidth]{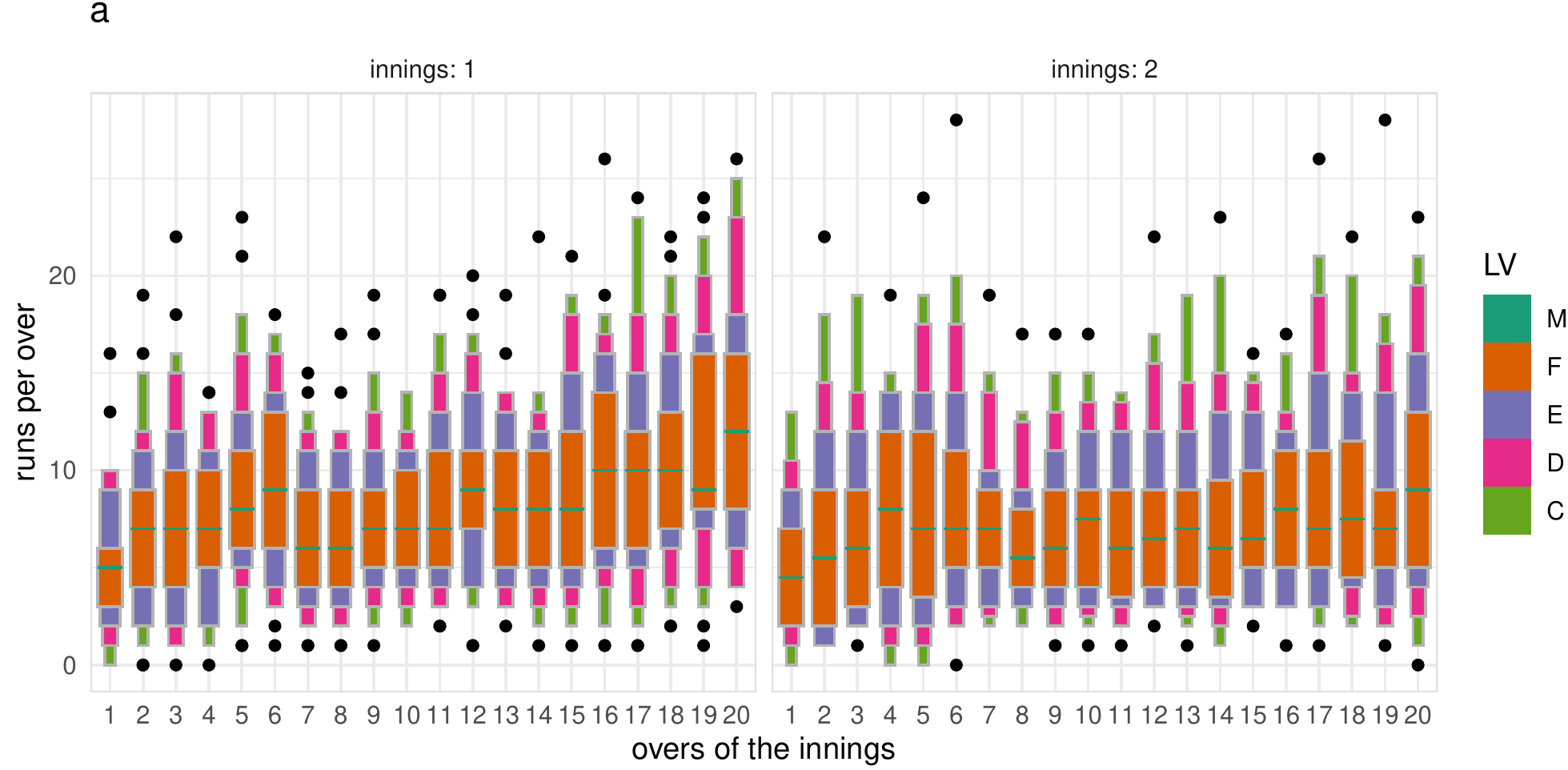} \includegraphics[width=\textwidth]{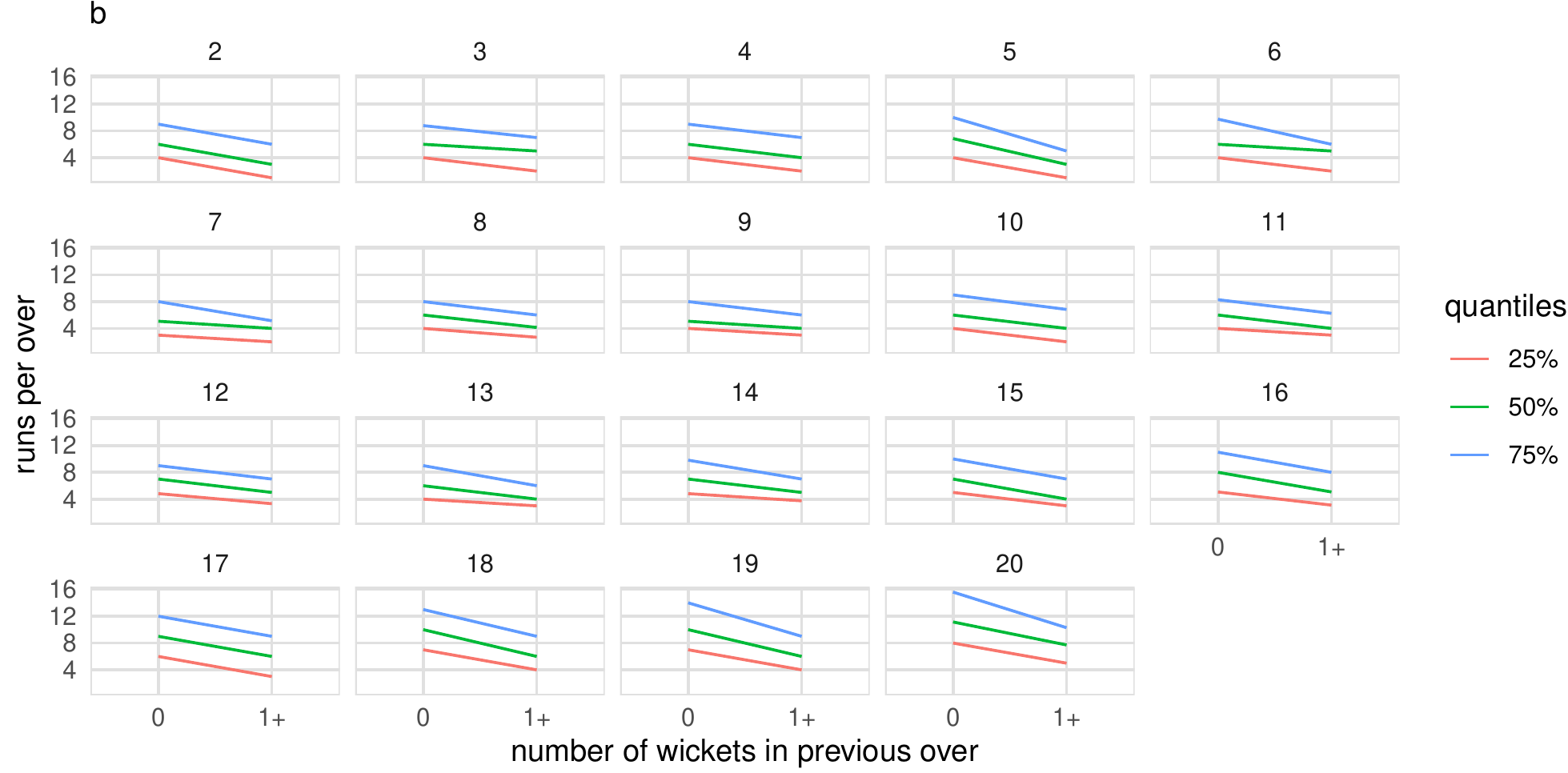} 

}

\caption{Runs per over shown with different distribution displays, and granularities. Plot a shows letter value plots across overs faceted by innings. For the team batting in the first innings there is an upward trend of runs per over, while there is no such pattern of runs for the teams batting in the second innings. Plot b shows quantile plots of runs per over across an indicator of wickets in the previous over faceted by current over. This indicates that at least one wicket in the previous over leads to lower median run rate and quartile spread in the subsequent over, regardless of the over of the innings.}\label{fig:cricex}
\end{figure}

There are many interesting questions that could be addressed with the \texttt{cricket} data set. For example, does the distribution of total runs vary depending on if a team bats in the first or second innings? The Mumbai Indians (MI) and Chennai Super Kings (CSK) appeared in final playoffs from 2010 to 2015. Using data from these two teams, it can be observed (\autoref{fig:cricex}a) that for the team batting in the first innings there is an upward trend of runs per over, while there is no clear upward trend in median and quartile deviation of runs for the team batting in the second innings after the first few overs. This suggests that players feel mounting pressure to score more runs as they approach the end of the first innings, while teams batting second have a set target in mind and are not subjected to such mounting pressure and therefore may adopt a more conservative run-scoring strategy.

Another question that can be addressed is if good fielding or bowling (defending) in the previous over affects the scoring rate in the subsequent over? To measure the defending quality, we use an indicator function on dismissals (1 if there was at least one wicket in the previous over, 0 otherwise). The scoring rate is measured by runs per over. \autoref{fig:cricex}b shows that no dismissals in the previous over leads to a higher median and quartile spread of runs per over compared to the case when there has been at least one dismissal in the previous over. This seems to be unaffected by the over of the innings (the facet variable). This might be because the new batsman needs to play himself in or the dismissals lead the (not-dismissed) batsman to adopt a more defensive play style. Run rates will also vary depending on which player is facing the next over and when the wicket falls in the previous over.

Here, wickets per over is an aperiodic cyclic granularity, so it does not appear in the hierarchy table. These are similar to holidays or special events in temporal data.

\hypertarget{sec:discussion}{%
\section{Discussion}\label{sec:discussion}}

Exploratory data analysis involve many iterations of finding and summarizing patterns. With temporal data available at ever finer scales, exploring periodicity can become overwhelming with so many possible granularities to explore. This work provides tools to classify and compute possible cyclic granularities from an ordered (usually temporal) index. We also provide a framework to systematically explore the distribution of a univariate variable conditional on two cyclic time granularities using visualizations based on the synergy and levels of the cyclic granularities.

The \texttt{gravitas} package provides very general tools to compute and manipulate cyclic granularities, and to generate plots displaying distributions conditional on those granularities.

A missing piece in the package \texttt{gravitas} is the computation of cyclic aperiodic granularities which would require computing aperiodic linear granularities first. A few R packages including \texttt{almanac}\citep{R-almanac} and \texttt{gs}\citep{R-gs} provide the tools to create recurring aperiodic events. These functions can be used with the \texttt{gravitas} package to accommodate aperiodic cyclic granularities.

We propose producing plots based on pairs of cyclic granularities that form harmonies rather than clashes or near-clashes. A future direction of work could be to further refine the selection of appropriate pairs of granularities by identifying those for which the differences between the displayed distributions is greatest, and rating these selected harmony pairs in order of importance for exploration.

\hypertarget{acknowledgments}{%
\section*{Acknowledgments}\label{acknowledgments}}
\addcontentsline{toc}{section}{Acknowledgments}

The Australian authors thank the ARC Centre of Excellence for Mathematical and Statistical Frontiers \href{https://acems.org.au/home}{(ACEMS)} for supporting this research. Thanks to \href{https://data61.csiro.au/}{Data61 CSIRO} for partially funding Sayani's research and Dr Peter Toscas for providing useful inputs on improving the analysis of the smart meter application. We would also like to thank Nicholas Spyrison for many useful discussions, sketching figures and feedback on the manuscript. The package \texttt{gravitas} was built during the \href{https://summerofcode.withgoogle.com/archive/}{Google Summer of Code, 2019}. More details about the package can be found at \href{https://sayani07.github.io/gravitas/}{sayani07.github.io/gravitas}. The Github repository, \href{https://github.com/Sayani07/paper-gravitas}{github.com/Sayani07/paper-gravitas}, contains all materials required to reproduce this article and the code is also available online in the supplemental materials. This article was created with \texttt{knitr} \citep[\citet{R-knitr}]{knitr2015} and \texttt{rmarkdown} \citep[\citet{R-rmarkdown}]{rmarkdown2018}.

\hypertarget{supplementary-materials}{%
\section{Supplementary Materials}\label{supplementary-materials}}

\textbf{Data and scripts:} Data sets and R code to reproduce all figures in this article (main.R).

\noindent
\textbf{R-package:} The ideas presented in this article have been implemented in the open-source R \citep{R-language} package \texttt{gravitas} \citep{R-gravitas}, available from CRAN. The R-package facilitates manipulation of single and multiple-order-up time granularities through cyclic calendar algebra, checks feasibility of creating plots or drawing inferences for any two cyclic granularities by providing list of harmonies and recommends possible visual summaries through factors described in the article. Version 0.1.3 of the package was used for the results presented in the article and is available on Github (\url{https://github.com/Sayani07/gravitas}).

\bibliographystyle{agsm}
\bibliography{bibliography.bib}

\end{document}